\documentclass[12pt,prd,aps,amssymb,amsmath,tightenlines]{revtex4}
\usepackage{psfig}

\usepackage{graphics}
\usepackage[pdftex]{graphicx}

\begin{document}
\preprint{}

\title{Bernoulli-like polynomials associated with Stirling Numbers}

\author{Carl M. Bender\footnote{Permanent address: Department of
Physics, Washington University, St. Louis MO 63130, USA}, Dorje C.
Brody, and Bernhard K. Meister}

\affiliation{Blackett Laboratory, Imperial College, London SW7
2BZ, UK}

\date{\today}

\begin{abstract}
The Stirling numbers of the first kind can be represented in terms
of a new class of polynomials that are closely related to the
Bernoulli polynomials. Recursion relations for these polynomials
are given.
\end{abstract}


\maketitle

The Stirling numbers have wide application in science and
engineering. For example, the generating function of the Stirling
numbers of the first kind appears in the inversion of a
Vandermonde matrix \cite{BBM}, which is used in curve fitting,
coding theory, and signal processing \cite{R}.

In this paper we show that there is a connection between the
Stirling numbers of the first kind and the Bernoulli polynomials.
The unsigned Stirling numbers of the first kind
$(-1)^{n-m}S_n^{(m)}$ are defined as the number of permutations of
$n$ symbols which have exactly $m$ permutation cycles \cite{C,AS}.
Let us construct the positive integers $T_{n,k}$ by
\begin{eqnarray}
T_{n,k}=(-1)^{k-1}S_n^{(n-k+1)}. \label{e1}
\end{eqnarray}
Here is a table of the numbers $T_{n,k}$:
\begin{eqnarray}
\begin{array}{ccccccccc}
T_{1,k}:&\quad&1&0&0&0&0&0&0\\
T_{2,k}:&\quad&1&1&0&0&0&0&0\\
T_{3,k}:&\quad&1&3&2&0&0&0&0\\
T_{4,k}:&\quad&1&6&11&6&0&0&0\\
T_{5,k}:&\quad&1&10&35&50&24&0&0\\
T_{6,k}:&\quad&1&15&85&225&274&120&0\\
T_{7,k}:&\quad&1&21&175&735&1624&1764&720
\end{array}
\nonumber
\end{eqnarray}
Note that the horizontal sums of the numbers in this table are
factorials. That is, $1+1=2!$, $1+3+2=3!$, $1+6+11+6 =4!$, and so
on.

Let us construct a new set of polynomials $P_k(n)$ that reproduces
these Stirling numbers. For example, the numbers
($0,~1,~3,~6,~10,~\ldots$) in the first column in this array as a
function of the row label $n$ are given by the polynomial
$$\frac{1}{2^1 1!}n(n-1)P_0(n),$$
where
\begin{eqnarray}
P_0(n)=1. \label{e3}
\end{eqnarray}
The numbers ($0,~0,~2,~11,~35,~\ldots$) in the next column in this
array as a function of $n$ are given by
$$\frac{1}{2^2 2!}n(n-1)(n-2)P_1(n),$$
where
\begin{eqnarray}
P_1(n)=n-\frac{1}{3}. \label{e4}
\end{eqnarray}
The numbers in the next two columns in this array are given by
$$\frac{1}{2^3 3!}n(n-1)(n-2)(n-3)P_2(n),$$
and
$$\frac{1}{2^4 4!}n(n-1)(n-2)(n-3)(n-4)P_3(n),$$
where
\begin{eqnarray}
P_2(n)=n^2-n, \label{e5}
\end{eqnarray}
and
\begin{eqnarray}
P_3(n)=n^3-2n^2+\frac{1}{3}n+\frac{2}{15}. \label{e6}
\end{eqnarray}
In general, the numbers along the $k$th column are expressible in
the form
$$\frac{1}{2^kk!}n(n-1)\cdots(n-k)P_{k-1}(n) .$$
The next seven polynomials $P_k(n)$ are
\begin{eqnarray}
P_4(n)&=&n^4-{\textstyle\frac{10}{3}}n^3+
{\textstyle\frac{5}{3}}n^2+{\textstyle\frac{2}{3}}n, \nonumber\\
P_5(n)&=&n^5-5n^4+5n^3+{\textstyle\frac{13}{9}}n^2-
{\textstyle\frac{2}{3}}n- {\textstyle\frac{16}{63}},\nonumber\\
P_6(n)&=&n^6-7n^5+{\textstyle\frac{35}{3}}n^4+
{\textstyle\frac{7}{9}}n^3- {\textstyle\frac{14}{3}} n^2-
{\textstyle\frac{16}{9}}n, \nonumber\\ P_7(n)&=&
n^7-{\textstyle\frac{28}{3}}n^6+{\textstyle\frac{70}{3}}
n^5-{\textstyle\frac{56}{9}}n^4- {\textstyle\frac{469}{27}} n^3
-4n^2+{\textstyle\frac{404}{135}}n+{\textstyle\frac{16}{15}},
\label{e7}\\
P_8(n)&=&n^8-12n^7+42n^6-{\textstyle\frac{448}{15}}n^5-
{\textstyle\frac{133}{3}}n^4+ {\textstyle\frac{20}{3}}n^3 +
{\textstyle\frac{404}{15}}n^2+ {\textstyle\frac{48}{5}}n,
\nonumber\\ P_9(n)&=& n^9-15n^8+70n^7 - {\textstyle\frac{266}{3}}
n^6-{\textstyle\frac{245}{3}}n^5+ {\textstyle\frac{745}{9}}n^4 +
{\textstyle\frac{1072}{9}}n^3+{\textstyle\frac{188}{9}}n^2-
{\textstyle\frac{208}{9}}n- {\textstyle\frac{256}{33}},
\nonumber\\ P_{10}(n)&=& n^{10}- {\textstyle\frac{55}{3}}
n^9+110n^8-{\textstyle\frac{638}{3}}n^7-
{\textstyle\frac{847}{9}}n^6 + {\textstyle\frac{3179}{9}}n^5+
{\textstyle\frac{968}{3}}n^4- {\textstyle\frac{1100}{9}}n^3-
{\textstyle\frac{2288}{9}}n^2 - {\textstyle\frac{256}{3}}n.
\nonumber
\end{eqnarray}

The structure of these monic polynomials strongly resembles that
of the Bernoulli polynomials $B_m(x)$ \cite{AS}. First, like the
Bernoulli polynomials, every other polynomial in (\ref{e3}) --
(\ref{e7}) has a constant term. Second, recall that the constant
terms in the Bernoulli polynomials are Bernoulli numbers,
$B_{2m}(0)= B_{2m}$, and observe that the constant terms in the
polynomials $P_{2m-1}(n)$ are expressible in terms of Bernoulli
numbers:
\begin{eqnarray}
P_{2m-1}(0)=-\frac{1}{2m}4^m B_{2m}. \label{e8}
\end{eqnarray}

Furthermore, both the polynomials $P_m(n)$ and the Bernoulli
polynomials $B_m(x )$ satisfy very similar recursion relations.
The polynomials $P_m(n)$ satisfy
\begin{eqnarray}
P_{2m}(n)=(2m+1)nP_{2m-1}(n)+\sum_{k=1}^m (-2)^k
\frac{m!(2m-k+1)}{(k+1)!(m-k)!}n^{k+1}P_{2m-k-1}(n), \label{e9}
\end{eqnarray}
while the Bernoulli polynomials obey the same recursion relation
with an inhomogeneous term:
\begin{eqnarray}
B_{2m+1}(x) &=& (2m+1)xB_{2m}(x)+\sum_{k=1}^m (-2)^k
\frac{m!(2m-k+1)}{(k+1)!(m-k)!}x^{k+1}B_{2m-k}(x) \nonumber \\ & &
\quad + {\textstyle\frac{1}{2}}(-1)^{m+1}x^{2m}. \label{e10}
\end{eqnarray}

Finally, we note that $P_m(n)$ satisfies a recursion relation in
terms of the coefficients of the Bernoulli polynomials:
\begin{eqnarray}
P_{m+1}(n)&=&nP_m(n)- \frac{n}{m+2} \sum_{k=1}^{[m/2]} 4^k
P_{m+1-2k}(n) \left[{\rm coefficient~of}~x^{m+2-2k}~{\rm
in}~B_{m+2}(x)\right] \nonumber\\  &&\qquad - 2^m
\left[\frac{4}{m+2}B_{m+2}(0)-2n B_{m+1}(0) \right]. \label{e11}
\end{eqnarray}

\vspace{0.5cm} DCB gratefully acknowledges financial support from
The Royal Society. CMB is supported in part by the U.S.~Department
of Energy.

\end{document}